\documentclass{elsart}

\usepackage{amssymb}
\usepackage{epsfig}
\newcommand{\be}{\begin{equation}}
\newcommand{\ee}{\end{equation}}
\newcommand{\ba}{\begin{array}}
\newcommand{\ea}{\end{array}}
\newcommand{\bea}{\begin{eqnarray}}
\newcommand{\eea}{\end{eqnarray}}

\begin{document}

\begin{frontmatter}






\title{Population Growth and Control\\ in Stochastic Models of Cancer Development}

\author{Anna Ochab-Marcinek\corauthref{cor_AOM}} and
\ead{ochab@th.if.uj.edu.pl}
\author{Ewa Gudowska-Nowak}
\ead{gudowska@th.if.uj.edu.pl}

\address{Marian~Smoluchowski Institute of Physics,\\
and Mark Kac Center for Complex Systems Research\\
Jagellonian University, Reymonta~4, PL-30--059~Krak\'ow, Poland}

\corauth[cor_AOM]{Corresponding author.}

\begin{abstract}
We study the joint effect of thermal bath fluctuations
and an external noise tuning activity of cytotoxic cells
on the triggered immune response in a growing  cancerous tissue.
The immune response is assumed to be primarily mediated by effector cells that
develop a cytotoxic activity against the abnormal tissue.
The kinetics of such a reaction is represented by an enzymatic-like
Michaelis-Menten
two step process. Effective free-energy surface for the process is further parameterised
by the fluctuating energy barrier between the states of high and low concentration of
cancerous cells. By analysing the far from equilibrium escape problem across the
fluctuating potential barrier, we determine conditions of the most efficient decay kinetics of
the cancer cell-population  in the presence of dichotomously fluctuating concentration of
cytotoxic cells.
\end{abstract}

\end{frontmatter}

\section{Introduction}
\label{Introduction}

Mathematical models of population growth based on nonlinear ordinary differential
equations have been widely studied (see e.g. \cite{Murray,Sachs_Hlatky}), since despite their simplicity they can often capture the essence of
complex biological interactions and explain characteristics of proliferation phenomena.
However, biological processes are not purely deterministic: systems existing
in nature are
 subject to different types of natural noises. Generally, one can divide them into two types:
(i)-internal noises (thermodynamic fluctuations) that display system parameters variability
 at the thermodynamic equilibrium,
 and (ii)-external (mostly conformational) noises, due solely to the effect of time-variability
 of the environment. In consequence, the information about dynamics of complex biological systems
 with underlying conformational transitions may be relevantly addressed by use
 of the mesoscopic approach in which the time-evolution of a collective
 coordinate is governed by  stochastic differential equations \cite{Lefever}. In such an
 approach, system kinetics is described in terms of an analogue to the
 Kramers scenario with the Brownian particle wandering over an effective free
 energy potential. In many cases, the potential experienced by the Brownian
 particle (or, correspondingly by a collective coordinate of the system) cannot
 be regarded as static but as affected by some random fluctuations with
 characteristic time scale that is comparable with one of the time scales
 governing the passage over the potential barrier separating basins of
 stationary states \cite{Doering_Gadoua,Zuercher_Doering,astumian,rei,iwa,gam}. Some examples include the escape of $\mathrm{O}_2$ or $\mathrm{CO}$ ligand
 molecule from the myoglobin after dissociation, the ion channel kinetics in the
 lipid cell membrane or protein folding \cite{astumian}.
 Another illustration  of a coupling between the external noise and
 and a collective variable is an effect of cell-mediated immune surveillance
  against the cancer \cite{GARAY}. Most of tumoral cells bear antigens which are recognised as
 strange by the immune system. A response against these antigens may be mediated either by immune
 cells such as T-lymphocytes or other cells, not directly related to the immune system (like
  macrophages or natural killer cells). The process of damage to tumour proceeds via infiltration
   of the latter by
  the specialised cells which subsequently develop a cytotoxic activity against the
  cancer cell-population.
  The series of cytotoxic reactions between the cytotoxic cells and the tumour tissue may be
  considered to be well approximated \cite{GARAY} by a saturating, enzymatic-like process
  whose time evolution equations are similar to the standard Michaelis-Menten  kinetics.
  The variability of kinetic parameters defining this process naturally affects
   the extinction of the
  tumour \cite{GARAY,Prigogine_Lefever}.\\

In this paper we investigate an extension of the above-mentioned problem: We discuss
 properties of a two state cancer-growth model
 subject to independent Markovian dichotomous noise (affecting the
 immunological response) and to an
 additive thermal
 noise (equilibrium fluctuations around stationary
 states representing low and high concentrations of cancer cells). As a model of the cancer cells dynamics, we have considered an
overdamped Brownian particle moving in a two-well quasi-potential
between reflecting and absorbing boundaries, in the presence of noise
 that modulates the height of the barrier dividing two stable states
  of the population. Transitions from one state to the other (here: from
  a fixed-size tumour to a cancer-free state or {\em vice versa \/}) are
  induced by additive thermal noise.
In models of this kind, it is of particular interest how fast the
 system escapes the potential well, surmounting the fluctuating potential barrier.
 In the case of the cancer growth model, we are especially interested
 in the rate at which the system escapes the fixed-size tumour state,
  that leads to the extinction of the tumour. The mean escape time from the meta-stable
  state across the fluctuating barrier may exhibit non-monotonic  dependence on the
  characteristic time scale of these fluctuations, as has been exemplified in other studies
  \cite{Doering_Gadoua,Zuercher_Doering,rei,iwa,Bier_Astumian,Reim,Reimann_Haenggi,Dybiec_Gudowska}.
  Here, the phenomenon is discussed for the model system in which existence of the
  dichotomous Markovian noise in one of the parameters may change relative stability of meta-stable
  states and therefore reverse the direction of the kinetic process.\\
The following section presents briefly a generic model system that has been used
for the analysis of cancer growth kinetics. Subsequent paragraphs are devoted to the
presentation of the deterministic model and analysis of its stochastic counterpart.

\section{Generic Model System}
\label{Generic_Model_System}
We adhere to the model of  an overdamped Brownian particle moving in a
 potential field between absorbing and reflecting boundaries in the
  presence of noise which modulates the barrier  height. The evolution of
  a state variable $x(t)$ is described in terms of the Langevin equation
\be
\frac{dx}{dt} \ = \ -\frac{dV(x)}{dx}+\sigma\xi(t)+g(x)\eta(t) \ = \ -\frac{dV^\pm(x)}{dx}+\sigma\xi(t).
\label{lang}
\ee
\noindent Here $\xi(t)$ is a Gaussian process with zero mean and correlation

\[
<\xi(t)\xi(t')>\ =\ \delta(t-t')
\]

\noindent \textit{i.e.} the Gaussian white noise of intensity
$\sigma=\sqrt{2k_BT}$
arising from the heat bath of temperature $T$), whereas $\eta(t)$ stands for a Markovian
dichotomous noise switching between two levels $\{\Delta^{+},\Delta^{-}\}$ with mean frequency $\gamma$ and correlation time $1/2\gamma$. This means that its autocorrelation function is

\[
\langle\ (\eta(t)-\langle\eta\rangle)\ (\eta(t')-\langle\eta\rangle)\ \rangle\ =\ \frac{\left(\Delta^+-\Delta^-\right)^2}{4}e^{-2\gamma|t-t'|}.
\]

\noindent Both noises are assumed to be statistically independent,
\textit{i.e.} $<\xi(t)\eta(s)>=0$.
Eq. (\ref{lang}) can be considered as describing an overdamped motion of the
 state variable, subject to  an effective force

\[
-\frac{dV^\pm(x)}{dx}=-\frac{dV(x)}{dx}+\Delta^\pm g(x).
\]

\noindent Based on eq. (\ref{lang}), one can write a set of the Fokker-Planck equations which describe the evolution of probability density of finding the state variable
 in a ``position'' $x$ at time $t$:

\begin{eqnarray}
\partial_t {p}(x,\Delta^\pm,t)&\  =\ &  \partial_x  \left[\ \frac{dV^\pm(x)}{dx}\ +\ \frac{1}{2}\sigma^2\partial_x \  \right]\ p(x,\Delta^\pm,t) \nonumber \\
  & &\  -\  \gamma p(x,\Delta^\pm,t)\ +\ \gamma p(x,\Delta^\mp,t)
\label{schmidr}
\end{eqnarray}

\noindent In the above equations time has dimension of $[length]^2/energy$ due to a
friction constant that has been ``absorbed'' in a time variable.\\
 With the initial condition

\be
p(V\pm,x_s,t)|_{t=0}\ =\ \frac{1}{2}\ \delta(x-x_s),
\ee

\noindent the equations for the mean-first-passage time ($MFPT$) read:

\bea
\left\{
\begin{array}{ll}
-\frac{1}{2}\  =\  -\gamma\tau^+(x)+\gamma\tau^-(x)-\frac{dV^+(x)}{dx}\frac{d\tau^+(x)}{dx}+\frac{\sigma^2}{2}\frac{d^2\tau^+(x)}{dx^2}&\\
-\frac{1}{2}\  =\  \gamma\tau^+(x)-\gamma\tau^-(x)-\frac{dV^-(x)}{dx}\frac{d\tau^-(x)}{dx}+\frac{\sigma^2}{2}\frac{d^2\tau^-(x)}{dx^2}&
\label{mfptbn}
\end{array}
\right.
\eea

\noindent where $\tau^+(x)$ and $\tau^-(x)$ denote $MFPT$ for $V^+(x)$ and $V^-(x)$, respectively.\\

\noindent The overall $MFPT$  for the system
\be
MFPT(x)=MFPT^+(x)+MFPT^-(x)=\tau^+(x)+\tau^-(x)
\ee

\noindent can be obtained after solving eq.(\ref{mfptbn}) with appropriate
 boundary conditions. Here we assume a
 motion between a reflecting boundary $a$ and an absorbing boundary $b$:

\be
\left\{
\begin{array}{ll}
\ \frac{d\tau^+(x)}{dx}|_{x=a}\ =\ 0&\\
\ \frac{d\tau^-(x)}{dx}|_{x=a}\ =\ 0&\\
\ \tau^+(x)|_{x=b}\ =\ 0&\\
\ \tau^-(x)|_{x=b}\ =\ 0&
\end{array}
\right.
\ee

\noindent According to a vast literature on the subject, the most interesting aspect
 of the escape  over such a fluctuating potential  is the non-monotonic behaviour
 of the $MFPT$ as a function of the driving noise
 \cite{Doering_Gadoua,iwa,Bier_Astumian,Reim,Reimann_Haenggi}.
 In particular,
we expect that for the frequency of potential switching $\gamma$ tending to
zero (i.e. in the limit of a long correlation time of the dichotomous noise
 $\eta(t)$) the overall $MFPT$ will be a mean value of $MFPT$s for both
 configurations,
\be \label{mfpt_0}
\lim_{\gamma\rightarrow 0}MFPT(V^+,V^-,\gamma)= \frac{1}{2}\left[MFPT(V^+)+MFPT(V^-)\right],
\ee
\noindent whereas
 for $\gamma$ tending to infinity (in the limit of a short correlation time)
  the system will ``experience'' a mean barrier:
\be
\lim_{\gamma\rightarrow\infty}MFPT(V^+,V^-,\gamma)=MFPT\left(\frac{V^+}{2}+\frac{V^-}{2}\right),\label{mfpt_infty}
\ee
\noindent with the $MFPT(V^+)$ and $MFPT(V^-)$ obtained from formula
\be
MFPT(x)=\frac{2}{\sigma^2}\int^b_xdy\exp\left(\frac{2V_{\pm}(y)}{\sigma^2}\right)\int_a^ydz
\exp\left(\frac{-2V_{\pm}(z)}{\sigma^2}\right)
\ee
for $V=V^\pm$, separately.\\
\noindent
Although the solution of~(\ref{mfptbn}) is usually unique,
 a closed, ``ready to use'' analytical formula for $\tau$ can be
  obtained only for the simplest cases of the potentials.
  More complex cases require either use of approximation
  schemes \cite{rei}, perturbative approach \cite{iwa} or direct numerical
   evaluation methods \cite{gam}.


\section{Population Model of Cancer Growth}

We use the predator-prey model (\cite{Lefever}, \cite{Prigogine_Lefever}) to
describe the cancer cells population growth in presence of cytotoxic cells.
The population dynamics can be described as follows: First, the cytotoxic
cells bind to the tumour cells at a rate proportional to the kinetic constant
$k_{1}$; second, the cancer cells which have been bound are killed and the
complex dissociates at a rate proportional to $k_{2}$.
The process can be described schematically:

\be
X\ +\ Y\ \longrightarrow \!\!\!\!\!\!\!\! ^{k_{1}} \ \ \ \ Z\ \longrightarrow \!\!\!\!\!\!\!\! ^{k_{2}}\ \ \ \  Y\ +\ P\ .
\label{scheme}
\ee
\noindent Here $X$ represents the population of tumour cells. Similarly, $Y$, $Z$ and
$P$ represent active cytotoxic cells, bound cells and dead tumour cells,
 respectively.
In a given (small) volume element, there is an upper limit $N$ to the number
of cells which may be present, given that each cell has a typical diameter
 equal to $a$. From now on, we will use normalised cellular densities:
  ($x=\frac{X}{N}$ instead of $X$, etc).
Following the original presentation \cite{Prigogine_Lefever}, we assume that
 (i) cancer cells undergo replication at a rate proportional
to the time constant $\lambda$; (ii) as a result of cellular replication
in limited volume, a diffusive propagation of cancer cells is possible, with
transport coefficient $\lambda a^{2}$; (iii) dead cancer cells undergo
 elimination at a rate proportional to a certain constant $k_{3}$; (iv)
  local cytotoxic cell population remains constant, i.e. $Y+Z=const$; (v)
   free cytotoxic cells can move with a ``diffusion'' coefficient $D$.
The spatio-temporal evolution of the tumour due to the above processes can
be then described by the set of balance equations:

\be\label{eq:sys1}
\left\{
\begin{array}{lll}
\frac{\partial x}{\partial t}&=&\lambda[1-(x+p)]x-k_{1}Eyx+ \lambda a^{2}(1-p)\nabla^2 x +\lambda a^{2}x\nabla^2 p \\

\frac{\partial y}{\partial t}&=&-k_{1}yx+k_{2}z+D\nabla^2 y \\
\frac{\partial z}{\partial t}&=&k_{1}yx-k_{2}z \\
\frac{\partial p}{\partial t}&=&k_{2}Ez-k_{3}p
\end{array}
\right.
\ee

\noindent where

\be
y+z=E=const.
\ee
In the limit when the effector cells diffuse much faster than the cancer cells
propagate by cellular replication and in which  the dead cells are rapidly
eliminated, the spatial distribution  of $Y$ and $Z$ cells equilibrates rapidly
with respect to the local density of living tumour cells and the above scheme of
kinetics can be  recasted in the form of the scalar problem
\cite{GARAY,Prigogine_Lefever}:

\be
\frac{\partial x}{\partial t}=(1-\theta x)x-\beta\frac{x}{x+1} + \nabla^2 x,
\label{model}
\ee

\noindent where
\[
 x=\frac{k_{1}x}{k_{2}},\ \theta=\frac{k_{2}}{k_{1}},\ \beta=\frac{k_{1}E}{\lambda},\ t=\lambda t.
\]
In our paper, we will consider the spatially homogeneous form 
of eq. (\ref{model}).\\
\noindent As presented here,  the model has a long history of analytical studies
(\cite{Prigogine_Lefever}, \cite{Banik}). In a
slightly modified form it turned out to be also of practical use
in biophysical modelling of radiation-induced damage production and processing
\cite{CUCINOTTA,KIEFER,SACHS}. In particular, the kinetic scheme eq.(\ref{scheme})
 has been adapted for the purpose of studying kinetics of
double-strand breaks rejoining and formation of simple chromosome exchange
aberrations \cite{CUCINOTTA} after DNA exposure to ionising radiation.
Similar kinetics has been proposed in analysis of saturable repair models
\cite{KIEFER,SACHS}
devoted to study evolution of radio-biological damage. In the latter,
the saturable
repair modelled by a Michaelis-Menten kinetics describes processing of damage
by enzyme systems that can be overloaded. \\
In the forthcoming section, we will analyse some versions of the model
eq.(\ref{model})
taking into account time variability of parameters as an effect of
the environmental noise.
 Below, following \cite{Prigogine_Lefever}, we briefly remind
 its deterministic properties.\\
 Equation (\ref{model}) can be considered as describing an overdamped motion of the state variable moving in a quasi - ``free energy potential'':

\be
V(x)=-\frac{x^2}{2}+\frac{\theta x^3}{3}+\beta x -\beta \ln(x+1).
\ee

\noindent where $V(x)$ has at most three extrema (stationary points of the system):

\be
x_1=0,
\ee

\be
x_2=\frac{1-\theta+\sqrt{(1+\theta)^{2}-4\beta \theta}}{2\theta}
\ee

\noindent and

\be
x_3=\frac{1-\theta-\sqrt{(1+\theta)^{2}-4\beta \theta}}{2\theta}.
\ee
Stability analysis reveals a strong dependence on $\theta$:

\noindent(i) For $\theta>1$ and $\beta>1$, $x_1$ is the only minimum of $V(x)$. For $\theta>1$ and $0<\beta<1$, $x_1$ becomes maximum and $x_2$ is a new minimum.

\noindent(ii) For $\theta<1$, outside the region
with $0<\beta<\frac{(1+\theta)^2}{4\theta}$, the properties of the system are
 identical to (i). Inside the region, $V(x)$ has two minima:  $x_1$, $x_2$ and
 one maximum at $x_3$.
For certain values of parameters, namely for
\be
\theta < 1,\ \ \ \ \  0 < \beta < \frac{(1+\theta)^2}{4\theta},
\ee
the system is bistable \footnote{Depending on $\theta$, there exists a unique
value of $\beta=\beta_0$, at which both states $x_1$ and $x_3$ are equally
stable. Further stochastic analysis of the model is performed for $\theta=0.1$,
and consequently $\beta_0\approx 2.669$}. In a forthcoming section, we will consider this system
subject to  the joint effect
 of independent noises:
to a multiplicative dichotomous noise in $\beta$ with exponential time-correlation
 and to a white
 additive noise representing thermal fluctuations.


\section{Stochastic Model of Cancer Growth}

\begin{figure}[t] 
\begin{center}
\epsfig{figure=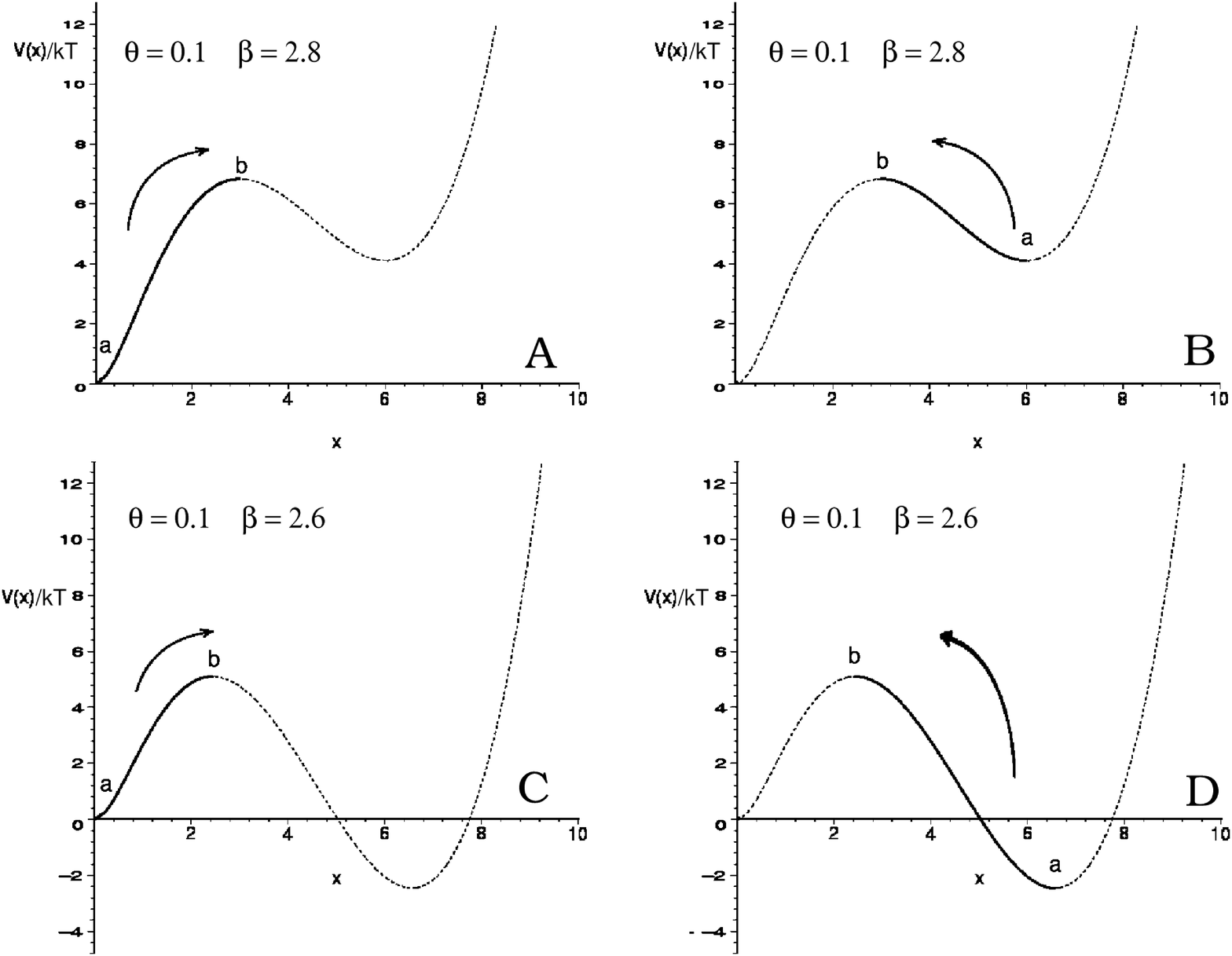, width=12cm}
\end{center}
\caption[]{\small Cancer growth kinetics has been studied for four indicated cases.
$MFPT$ was computed for intervals $[a,b]$, where $a$ and $b$ are reflecting and
 absorbing boundaries, respectively. Arrows point directions of a transition.}
\label{szkice}
\end{figure}

We investigate the system described by the equation
\be
\frac{dx}{dt}=(1-\theta x)x - \big(\beta+\Delta^\pm\big)\frac{x}{x+1}+\sigma \xi(t), \label{system}
\ee

\noindent where $\Delta^\pm$ is a two-state, Markovian  noise and  $\xi(t)$
is the Gaussian noise of intensity $\sigma$. In this form the model includes
the influence of the heat bath (modelled by a memoryless Gaussian noise) and
 the fluctuations in the
immunological response of the organism (here assumed to be represented by
a symmetric dichotomous noise in parameter $\beta$). Correspondingly, the process of population growth and
decay can be described as a motion of a fictitious particle in a
  potential switching between two conformational states (cf. Fig. \ref{szkice}).

\be
V^\pm(x)=-\frac{x^2}{2}+\frac{\theta x^3}{3}+(\beta \pm \Delta) (x - \ln(x+1)).
\ee


\begin{figure}[t] 
\begin{center}
\epsfig{figure=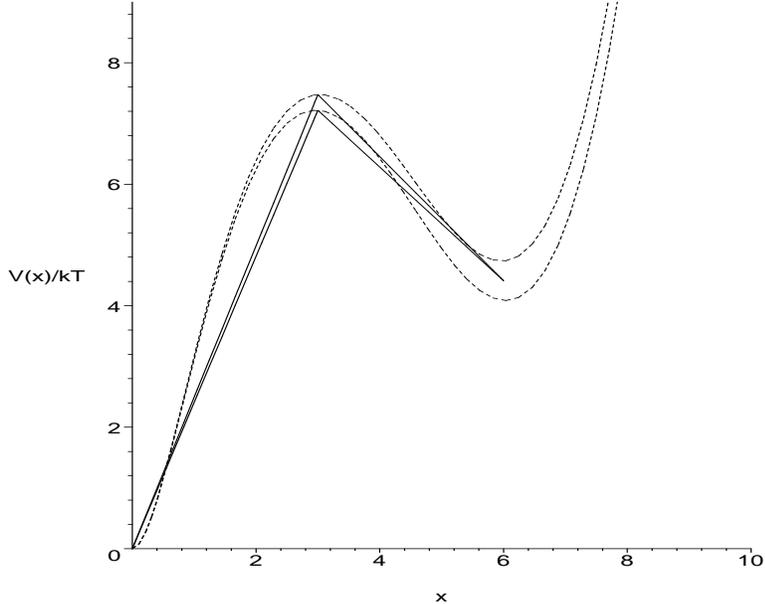, width=10cm, height=8cm}
\end{center}
\caption[]{\small Linear approximation of the original potential. The line connects minima of the potential $V(x,
\Delta=0)$ with the maximum of $V(x,\Delta^\pm)$. Dotted lines: 
$V(x,\Delta^\pm)$, solid lines: the approximated potential. }
\label{lin}
\end{figure}

For negligible additive noise and small concentration of cancerous cells, this
model resembles standard Verhulst equation with a perturbing multiplicative
dichotomous noise and as such has been shown \cite{Lefever} to exhibit a
complex scenario of noise-induced transitions observable in a pattern of the
stationary probability density. Here, we will address kinetic properties of this
model by studying the mean first passage time (\ref{mfptbn}) between high and low
population states in the system. In order to compute $MFPT$, as a working
example we use an approximation of the potential by a linear slope with a reflecting
 barrier placed at  $x_1=a=0$ and and absorbing barrier at $x_2=
 b=[1-\theta+\sqrt{(1+\theta)^2-4\beta\theta}](2\theta)^{-1}$. In a linear
 setup, it is frequently possible to formulate analytical solutions to the
 problem \cite{Zuercher_Doering,Bier_Astumian}. Moreover, as expected based on
 former analysis \cite{Dybiec_Gudowska}, due to the smoothing of the kinetics by
 irregular diffusive motions, the solutions for the piecewise linear potential
 are not qualitatively different from those for smooth differentiable potentials.
 In such a case a fully analytic expression for the $MFPT$  can be obtained,
 even though the algebra involved in such an evaluation requires use of
 symbolic computer softwares (here: Maple 7 procedure
 \texttt{dsolve/numeric/BVP}). \\
Based on the original potential, we build its linear approximation in
the following way: In general, inclusion of a noise term in the parameter $\beta$
influences positions of stationary states of the potential $V(x, \Delta^\pm)$
and affects their relative stability. However, if the effective change in the
barrier height is very small compared to its total height, the location of
extrema can be considered constant (cf. Fig.(\ref{lin})). Our analysis
demonstrates that even in such cases of weak perturbation of the barrier height,
presence of noise has dramatic consequences on the overall kinetics of cancer
and its extinction.

In order to keep constant the positions
 $x_1$, $x_2$ and $x_3$ of the extrema when changing the barrier height
  $V(x, \Delta^\pm)$, we approximate them with positions of the extrema
  of the potential $V(x,\Delta=0)$. Thus,\
   $x_1=0$, $x_2=\frac{1-\theta+\sqrt{(1+\theta)^{2}
   -4\beta \theta}}{2\theta}$ and $x_3=\frac{1-\theta-\sqrt{(1+\theta)
   ^{2}-4\beta \theta}}{2\theta}$.
Next, we connect points $[x_1,V(x_1, \Delta=0)]$, $[x_2,V(x_2, \Delta^+)]$
and $[x_3,V(x_3, \Delta=0)]$ or, respectively, $[x_1,V(x_1, \Delta=0)]$,
 $[x_2,V(x_2, \Delta^-)]$ and $[x_3,V(x_3, \Delta=0)]$ with straight lines, as
 has been depicted in Fig.\ref{lin}.\\
By defining
\be
A_\Delta:=\sqrt{1 + \theta ^{2} + 2\,\theta \,(1 - 2\,(\beta+\Delta ))}
\ee
and
\be
A_0 := A_{ \Delta=0}=\sqrt{1 + \theta ^{2} + 2\,\theta \,(1 - 2\,\beta)},
\ee
the height of the barrier seen by the particle located close to the $x_3$
can be expressed as
\[
V\left(x_3(\Delta),\Delta\right)\ -\  V\left(x_2(\Delta),\Delta\right) \ = \  \frac{-A_\Delta(9 \ \theta^2-12 \ \theta+3+A_\Delta^2)}{12 \ \theta^3}\ +
\]
\be
 +\  (\beta+\Delta)\left( \frac{-A_\Delta}{\theta} - \ln\left(\frac{\theta+1-A_\Delta} {\theta+1+A_\Delta}\right)\right)
 \label{pierw}
\ee
\noindent Similarly, the height of the barrier calculated from the bottom of
the left-side minimum $x_1=0$ is
\[
V\left(x_3(\Delta),\Delta\right) \ = \ \frac{(\theta-1+A_\Delta)^2 (4\theta-1+A_\Delta)}{24\theta^3} \ +
\]
\be
+ \ (\beta+\Delta)\left(\frac{-\theta+1-A_\Delta}{2\theta}-\ln\left(\frac{\theta+1-A_
\Delta}{2\theta}\right)\right)
\ee

Piecewise linear approximation of the potential results in:
\[
V \left(x_3(0),\Delta\right) \ - \ V\left(x_2(0),\Delta\right) \ = \  -\frac{A_0(-12\theta+9\theta^2+A_0^2+3)}{12\theta^3} \ +
\]
\be
+\beta\left(\frac{-A_0}{\theta}-\ln\left(\frac{\theta+1-A_0}{\theta+1+A_0}\right)\right)
+ \Delta\left(\frac{-\theta+1-A_0}{2\theta}-\ln\left(\frac{\theta+1-A_0}{2\theta}\right)\right)
\ee
and \\

\[
V\left(x_3(0),\Delta\right) \ = \ -\frac{(\theta-1+A_0)^2(4\theta-1+A_0)}{24\theta^3} \ +\]
\be
 + \ (\beta+\Delta) \left(\frac{-\theta+1-A_0}{2\theta}-\ln\left(\frac{\theta+1-A_0}{2\theta}\right)\right).
\label{ost}
\ee

\begin{figure}[t] 
\begin{center}
\epsfig{file=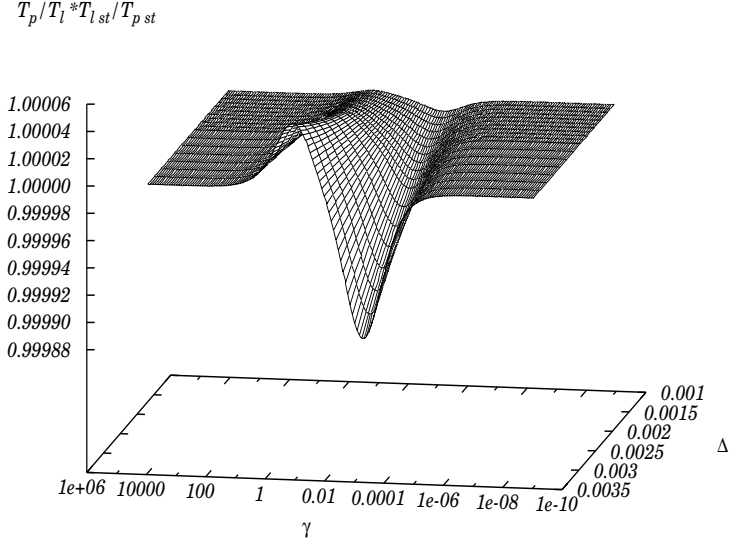, height=12.0truecm, width=10truecm,angle=-90}
\end{center}
\caption[]{\small Relative $MFPTs$ as a function of the rate of the barrier
fluctuations $\gamma$ computed for various values of $\Delta$, for an
approximated
linear potential. Parameters of the generic model has been set to
$\beta=2.669=\beta_0$, $\sigma=0.5$ and $\theta=0.1$.}
\label{3d}
\end{figure}

\noindent For the purpose of analysis,
the $MFPT$ has been computed for 4 chosen cases (cf. Figure 1):
(i) transition from large to small population when the small population state is a global minimum;
(ii) transition from large to small population when the large population state is a global minimum;
(iii) transition from small to large population when the small population state is a global minimum.
(iv) transition from small to large population when the large population state
 is a global minimum.\\
Figure \ref{3d} displays the graphs of $MFPT$ as a function
of the barrier fluctuation rate $\gamma$ and the noise intensity $\Delta$.
The results are presented in the form of a ratio $T_p/T_l$ of forward/backward
mean first transition times calculated as a fraction of a corresponding ratio
of transition times estimated for static ($\Delta=0$) barriers.\\

The results indicate that the escape kinetics
 (from right to left and {\it vice versa}) depends on the
slope of the approximated linear potential ({\it cf.}
eqs.(\ref{pierw}-\ref{ost}) and exhibits characteristic features
of the noise-induced
enhancement of the activation process.
Moreover, for the critical (deterministic) value of parameter $\beta=\beta_0$, 
fluctuations of the barrier may facilitate the forward/backward
transfer and within given intervals of frequencies either the growth or
diminishment of cancer population can be expected.
For either of the transitions (from small- to large-  and from large- to small 
concentrations of cancerous
cells),  
depending on the frequency of the barrier fluctuations $\gamma$, there exists
 an optimal value of the barrier
fluctuation rate, for which
the mean first passage time is minimal
\cite{Doering_Gadoua,Zuercher_Doering,iwa,Reimann_Haenggi,Dybiec_Gudowska}). The analysis reveals that
the intensity of the effect depends on the strength of the barrier noise
(the minimum/maximum of the relative $T_p/T_l$ deepens/lowers for larger
$\Delta$). Also, the resonant frequency  of the $MFPT$ minimum $\gamma_{min}$
increases monotonically with the intensity $\sigma$ of the additive
white Gaussian noise as displayed in Figure \ref{gvs}.

\begin{figure}[t] 

\begin{center}

\epsfig{file=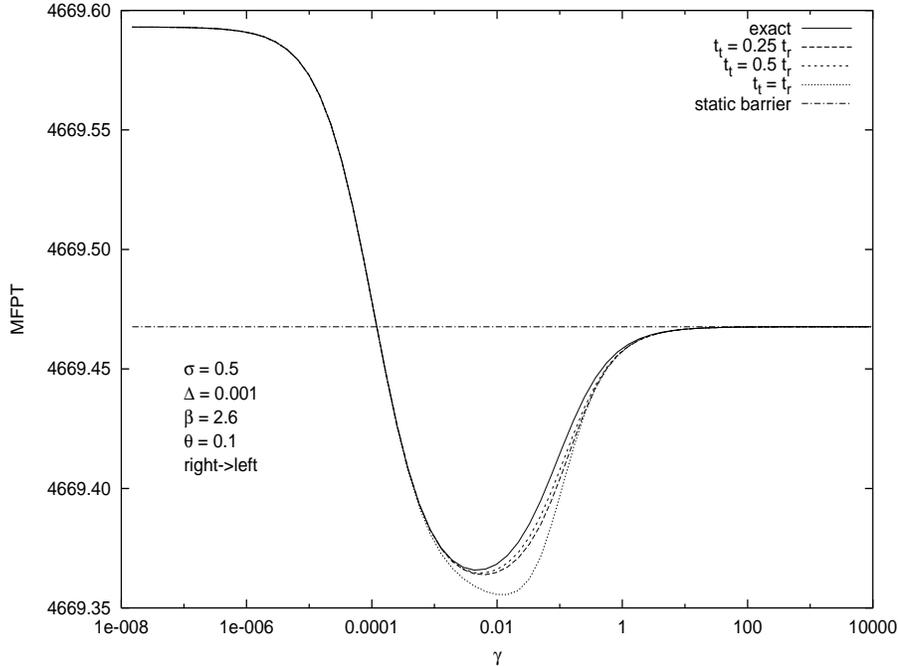, height=12.0truecm, width=9truecm,angle=-90}

\end{center}

\caption[]{\small Relative escape time as a function of the rate of the barrier
fluctuations $\gamma$ estimated for various duration of the integration time
$t_t$ as compared with the down-hill relaxation time $t_r$ for a static barrier.}
\label{wyk06}

\end{figure}
\begin{figure}[h] 

\begin{center}

\epsfig{file=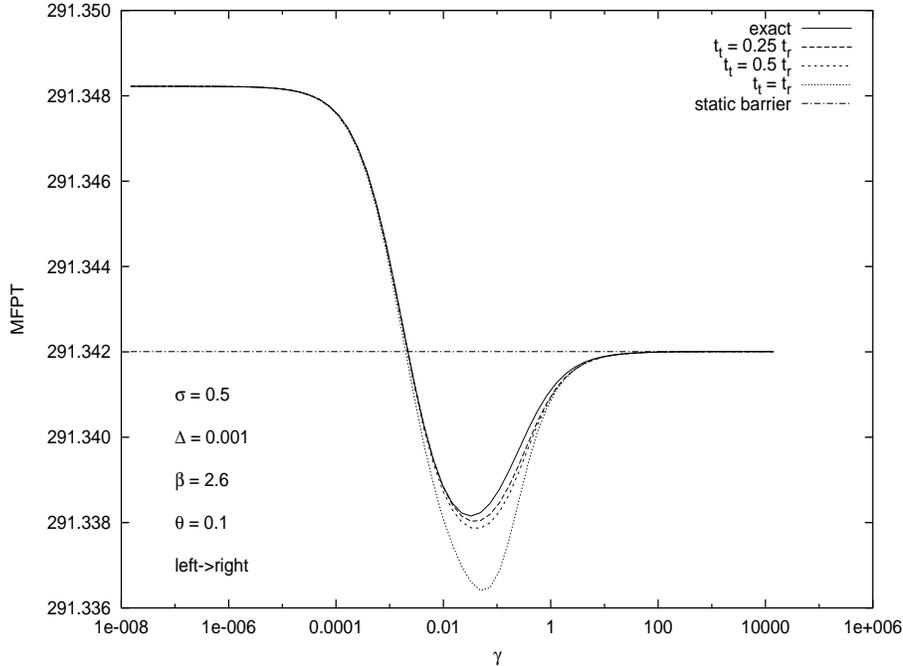, height=12.0truecm, width=9truecm,angle=-90}

\end{center}

\caption[]{Same as in Figure \ref{wyk06}, reverse transition over a fluctuating
barrier.} 
\label{wyk07}
\end{figure}


Numerically estimated $MFPT$ values have been compared with their asymptotes for $\gamma\rightarrow\infty$, $\gamma\rightarrow 0$ evaluated directly from formulas (\ref{mfpt_infty}), where
\be
MFPT(x) = \frac{x-b}{A}+\frac{\sigma^2}{2A^2}[\exp{\frac{2A}{\sigma^2}(b-a)}-\exp{\frac{2A}{\sigma^2}(x-a)}]
\ee
and
\be
A=\frac{V(a)-V^\pm(b)}{a-b}
\ee
for $MFPT(V^\pm)$, or
\be
A=\frac{V(a)-V(b)}{a-b}
\ee
for $MFPT(\frac{V^++V^-}{2})$, where $a=x_1$ and $b=x_2$.
 The results display full agreement with numerical evaluation.

Finally, we have used a recently proposed \cite{iwa}  approximated method for studying activation over a
fluctuating barrier that involves considering separately the slow and fast
components of barrier fluctuations and applies for any value of the correlation
time $\tau=1/\gamma$. By decomposing the barrier noise into two independent
terms
\bea
\eta(t)=\eta_s(t)+\eta_f(t)
\eea
the escape problem may be considered as a three-dimensional Markovian process
described by a joint probability distribution $P(x, \eta_s,\eta_f,t)$  which
evolves accordingly to the Fokker-Planck equation with two separate operators
responsible for the time evolution of slow and fast variables. In particular, by definition,
 $\eta_s(t)$ remains constant while the trajectory climbs the barrier and its
 dynamics can be analysed by the kinetic approach \cite{iwa,Bier_Astumian}. On the
 contrary, $\eta_f$ vanishes for the barrier correlation times slightly greater
 than the relaxation time from the top of the barrier to the bottom of
 the well, so that the probability distribution $P(x, \eta_s,\eta_f,t)$ can be
 deconvoluted to the form $p(x,\eta_f,t;\eta_s)\varrho(\eta_s,t)$ and in
 consequence, the
  rate theory formalism  can be safely applied. The equilibration
 process in a fast $\eta_f(t)$ variable leads then to an effective Fokker-Planck
 equation (with a new form of the quasi-potential \cite{iwa,Reim})
 from which the appropriate effective $MFPT(\eta_s)$ can be calculated. For
 dichotomous switching the procedure captures ideology of the kinetic rate
 estimated as an inverse of
 \bea
 MFPT(\eta_s)=\frac{k^++k^-+4\gamma}{2[k^+k^-+(k^++k^-)\gamma]},
 \eea
 where kinetic rates $k^+$, $k^-$ describe escape kinetics in two different
 configurations of the barrier. Note, that although the above formula resembles
 a typical kinetic scenario \cite{Bier_Astumian} which is known not to produce
 the resonant activation effect, the evaluation of the $MFPT(\eta_s)$ takes into
 account fast kinetics of the dynamics hidden in the form of kinetic constants
 evaluated from the effective Fokker-Planck equation. Crucial for the approach
 is the determination of the value of the integration interval $t_t$ which is
 describing climbing stage of the process. 
 Its duration for
 the unperturbed potential equals the relaxation time $t_r$ from the top to the
 bottom of the well. Obviously, fluctuations of the potential break that
 equality \cite{Bier_Astumian,iwa}. However, since we are not discussing the relationship
 between the process of climbing up and relaxing down the fluctuating barrier,
 but rather need a time-estimate for the following escape event, as a first
 approximation for $t_t$ the $MFPT$ from the top of the barrier to the bottom 
 of the well has been used.\\ 
\begin{figure}[t] 
\begin{center}
\epsfig {figure=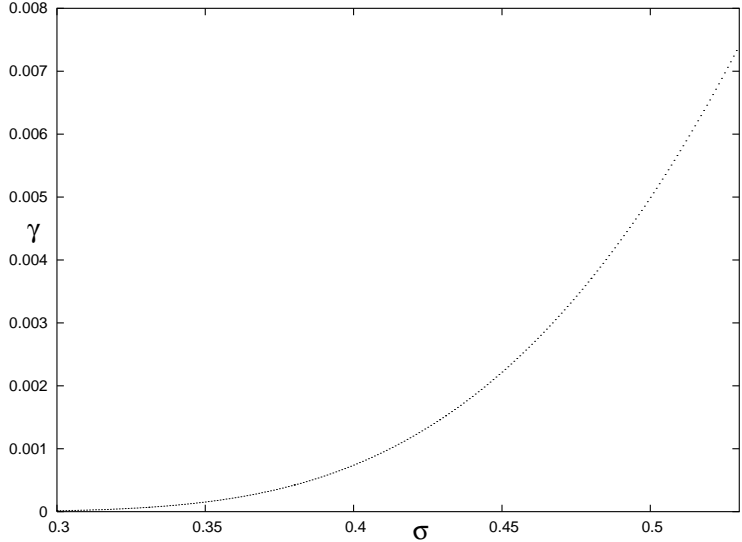, angle=270, width=10.5cm}
\end{center}
\caption[]{Position of the MFPT minimum \small $\gamma_{min}$  as a function of noise intensity $\sigma$ \\
}
\label{gvs}
\end{figure}

 To test the method, we have presented in Figures \ref{wyk06} and
 \ref{wyk07} the exact analytical results for the triangular barrier
 ({\it cf.} Figure \ref{lin}) along with the approximation based on the
 approach described above. The agreement with the original results is quite good
 and depends only slightly on the time of integration $t_t$ of the slow
 $\eta(t)$ component, as discussed
 elsewhere \cite{iwa}. Nevertheless, we expect that such a comparison will be less
 positive for kinetics over a relatively low barriers (data not shown).


\section{Conclusions}

We have considered a standard problem of an escape over a fluctuating potential
barrier for a system describing efficiency of the immunological response against
cancer. The basic component of the model is the mode of action of
immunologically active cells that behave as catalysts in a chemical reaction.
  Analytical results for the model have been obtained after approximating
a smooth double well potential by a piecewise-linear one. Although the method
presented in this paper allows to estimate only the shortest possible time of
transition (by placing an absorbing boundary condition at the top of the
barrier), the results demonstrate fine sensitivity of the process to the
correlation time of the barrier noise. In particular, by controlling the
frequency of barrier fluctuations (or equivalently, the correlation time of
changes in the effector cells response), the process of tissue growth can be
reversed. Further analysis should tackle the problem of obvious nonlinearity of
the process and the interplay of the noise with the stability criteria for the
double-well system. The formalism of a partial noise-averaging method \cite{iwa}
may provide a way to analyse this case.\\
It seems worthy mentioning that despite its simplicity, the model eq.(11) can be 
considered  as a generic one for kinetic description of many processes 
that  use allosteric transition and phosphorylation of proteins in their
metabolic pathways \cite{CUCINOTTA,KIEFER}. However, to ensure modelling of relevant
interactions and to determine proper estimates of the kinetic parameters, an
additional computer analysis of rate constants is required, which in turn has to
rely on experimental observations. In particular, local stability analysis of
such models leads to conditions on rate parameters and steady state 
concentrations that need to be compared with experimental data.   
If the binding kinetics was controlable in such cases by an external (stochastic
or deterministic) time-dependent
field, the analysis similar to presented in this paper could be used for
 prediction of a most effective  action of
the catalyst leading to an expected shortest reaction time.
 In modelling tumor treatment
or experiments on cell killing by external agents based on ordinary differential
equations, a very common assumption is
that treatment modifies the growth, so that the Verhulst- or Gompertz-
type growth rate has to be accompanied by an extra term supressing the
proliferating population \cite{SACHS}. The inhibition and deceleration of the
tumor growth can  often be assumed to follow a saturable repair model
\cite{Sachs_Hlatky,CUCINOTTA,KIEFER,SACHS} of the Michaelis-Menten kinetics. Therefore,
consequences of the above analysis may appear realistic in those situations
where, given a full identification of the parameters, the cellular populations
would exhibit extinction controlled by external, time dependent fields like
time-dependent administration of cytotoxic agents or therapeutic radiation.

\begin{ack}
We would like to express our gratitude to Dr. Jan Iwaniszewski for helpful and inspiring discussions.
\end{ack}



\end{document}